\documentclass[twocolumn,aps,showpacs,preprintnumbers,superscriptaddress]{revtex4}
\usepackage{graphicx}
\usepackage{dcolumn}
\usepackage{bm}

\begin{document}
\title{Phase diagram to design passive scatterers}
\author{Jeng Yi Lee}
\affiliation{
Institute of Photonics Technologies, National Tsing-Hua University, Hsinchu 300, Taiwan}

\author{Ray-Kuang Lee}
\affiliation{
Institute of Photonics Technologies, National Tsing-Hua University, Hsinchu 300, Taiwan}
\affiliation{Physics Division, National Center of Theoretical Science, Hsinchu 300, Taiwan}
\date{\today}

\begin{abstract}
A phase diagram, defined by the amplitude square and phase of scattering coefficients for  absorption cross-section in each individual channel,  is introduced as a universal map on the electromagnetic properties for passive scatterers. 
General physical bounds are naturally revealed based on the intrinsic power conservation in a passive scattering system, entailing  power competitions among scattering, absorption, and extinction.
Exotic scattering and absorption phenomena, from resonant scattering, invisible cloaking, coherent perfect absorber, and subwavelength superscattering can all be illustrated in this phase diagram. 
With electrically small core-shell scatterers as an example, we demonstrate a systematic method to design field-controllable structures based on the allowed trajectories in the phase diagram. 
The proposed phase diagram not only provides a simple tool to design optical devices but also promotes a deep understanding on Mie's scattering theory.
\end{abstract}
\pacs{42.25.Bs, 81.05.Zx, 42.25.Fx, 78.67.Bf}

\maketitle

Even though studies on scattering of light by small particles can been tracked back to Mie's seminal work one century ago ~\cite{Mie}, it was realized in the past decade to have an efficient way in manipulating and designing electromagnetic properties of light with nanostructures~\cite{nano}. 
In particular, unusual scattering characteristics,  such as resonant scattering~\cite{resonance-1, resonance-2,resonance-3,resonance-4}, coherent perfect absorption~\cite{perfect1,perfect2,perfect3,perfect4}, invisible cloaking~\cite{cloak1,cloak2,cloak3,cloak4,cloak5,cloak6}, subwavelength superscattering~\cite{superresonance1,superresonance2,superresonance4}, and minimum-scattering superabsorbers ~\cite{superabsorber1,superabsorber2}, are revealed in a single, isotropic, and homogeneous scatterer by embedding multi-layered structures.
Combined with state-of-the-art fabrication technologies, these meta-structures sever as   functional nano-devices with promising applications in light harvesting ~\cite{solar1,solar2,solar3}, heat generation by metal nanoparticles ~\cite{med1,med2}, optical nanocircuits ~\cite{circuit1}, and nonlinear optical processes ~\cite{nonlinear1,nonlinear2}.

To have exotic scattering properties, a variety of specific conditions are asked to be satisfied for the desired cross-sections constituted by scattering coefficients.
Undoubtedly, a better understanding in the scattering coefficients provides an access to design nanostructures, such as information about the limitation, energy assignment,  and robustness on the corresponding extrinsic field response of real scatterers.
For the working frequency of interests, for example, in the visible spectra many metals possess negative permittivities associated with a strong dispersion, which introduces real loss effect and suppresses the desired functions ~\cite{superresonance3}.
As possible mismatching in physical parameters  may happen, it is naturally to seek optimized  invisible cloaks or performance boundary in a cloaked sensor with  the consideration of intrinsic loss in reality  ~\cite{cloaksensors1}.
 
In this Letter, we study the general relation between the amplitude and phase in the scattering coefficients for a passive spherical scatterer under any frequency excitation.
By decomposing absorption powers for each individual channel, a phase diagram is introduced by recasting absorption cross-section in terms of the amplitude square and phase of scattering coefficients, which acts as a universal map to design passive scatterers.
Not only  all the physically allowed regions can be defined to satisfy the intrinsic power conservation, but also all the exotic electromagnetic properties in the literature can be illustrated in this phase diagram.
Moreover, as a demonstration, we take electrically small core-shell scatterers as an example to illustrate a systematic way in designing the composition of subwavelength-structures with required scattering and absorption properties.
Our work not only  promotes an alternative interpretation  of Mie's scattering theory, but also provides a compact solution to realize a variety of functionally subwavelength-devices.

We consider a linearly polarized plane wave with time evolution $e^{-i\omega t}$ at the frequency $\omega$, which is illuminating on a single spherical object. 
The object could be  made by multiple layers of uniform and isotropic media with complex permittivity and permeability.
The real and imaginary parts of permittivity (permeability) are denoted as  $\epsilon=\epsilon^{\prime}+i\epsilon^{\prime\prime}$ ($\mu=\mu^{\prime}+i\mu^{\prime\prime}$), where $\epsilon^{\prime\prime}$ ($\mu^{\prime\prime}$) is assumed to be a positive real number for a passive medium.
Without loss of generality, the surrounding environment is taken as non-absorptive, non-magnetic, and free of external sources or currents, i.e., $\epsilon_{0}=1$ and $\mu_{0}=1$.
Due to the free divergence in the governing Maxwell's equation, the electric field $\vec{E}$ and magnetic field $\vec{H}$ in the environment can be expressed by two auxiliary vector potentials, i.e., the transverse magnetic (TM) and transverse electric (TE) modes, which  are respectively generated by  two scalar  spherical  wave equations ~\cite{resonance-3,book1, book2}.
Each scalar function can be built by an infinite series with unknown coefficients determined through the  boundary conditions.
By following the conventional notations, let the  scattering coefficients be $S_{n}^{TM}$ and $S_{n}^{TE}$ for TM and TE modes in each spherical harmonic channel labeled by the index $n$, respectively.
It is known that the corresponding absorption and scattering cross-sections, $\sigma^{\text{abs}}$ and $\sigma^{\text{scat}}$, defined as the total power absorbed and scattered by a single scatterer with respect to the unit intensity of incident plane wave, can be expressed as
\begin{eqnarray}
&&\sigma^{\text{abs}} \equiv \sum_{n=1}^{\infty} \sigma^{\text{abs}(\text{TE})}_{n} + \sigma^{\text{abs}(\text{TM})}_{n} =  \\
&&-\sum_{n=1}^{\infty} \frac{(2n+1)\lambda^2}{2\pi} ( \text{Re}\{ S_{n}^{\text{TM}}+S_{n}^{\text{TE}}\}  +\vert S_{n}^{\text{TM}}\vert^{2}+\vert S_{n}^{\text{TE}}\vert^{2} ), \nonumber \\
&&\sigma^{\text{scat}}=\sum_{n=1}^{n=\infty}\frac{(2n+1)\lambda^2}{2\pi}( \vert S_{n}^{\text{TM}}\vert^{2}+\vert S_{n}^{\text{TE}}\vert^{2} ),
\end{eqnarray}
where $\lambda$ is the wavelength of incident wave in vacuum ~\cite{book1, book2}. 
Due to the conservation of power, the absorption cross-section is required to be equal or larger than zero for each channel, i.e., $\sigma^{abs(\text{TE},\text{TM})}_{n} \equiv - \frac{\lambda^{2}}{2\pi}(2n+1)(\text{Re}\{S_{n}^{(\text{TE},\text{TM})}\}+\vert S_{n}^{(\text{TE},\text{TM})}\vert^{2})  \ge 0$. 
Moreover, for a given radius of particle, denoted as $a$, the value of  size parameter $2\pi a/\lambda$ determines how many terms of this  convergent series to be dominant ~\cite{book2}.

In general, these scattering coefficients  control the extrinsic states in the scattering  field pattern.
Once the composition of scatterer  and geometry are given, the corresponding scattering coefficients are determined. 
Conversely, without knowing the composition in a scatterer, one can also have the same scattering coefficients.
In the following, we introduce a systematic way to do the {\it inverse design} for the scatterer by specifying the required scattering or absorption properties first.
To do that, we introduce the phase diagram by  expressing the scattering coefficient as $S_{n}^{(\text{TE}, \text{TM})}=\vert S_{n}^{(\text{TE}, \text{TM})} \vert \exp\{i\theta_{n}^{(\text{TE}, \text{TM})}\}$, where the magnitude $\vert S_{n}^{(\text{TE}, \text{TM})}\vert$ is a positive real value for every individual channel and $\theta_{n}^{(\text{TE}, \text{TM})}$ is the corresponding phase shift.
Since TM and TE modes are separable, we simplify the notations by neglecting the superscripts, i.e.,  re-writing the amplitude square and phase of scattering coefficients as $\vert S_{n}\vert^2$ and $\theta_{n}$.
In Fig. 1, by plotting the absorption cross-section of individual channel, normalized  to  the common factor $2\pi/(2n+1)\lambda^2$,  as a function of phase and amplitude square of scattering coefficient ($\theta_n, \vert S_n\vert^2$), we introduce the universal phase diagram for a passive scatterer.

\begin{figure}[t]
\centering
\includegraphics[width=8.2cm]{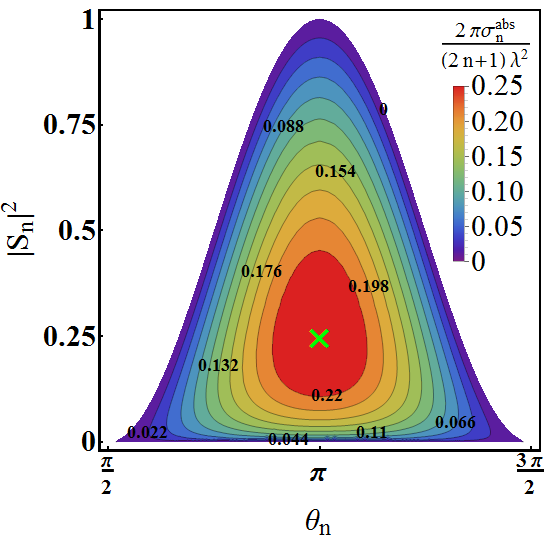}
\caption{(Color online) Phase diagram for each spherical harmonic channel is generated  in terms of the phase $\theta_n$ and amplitude square $|S_n|^2$ for the corresponding normalized absorption cross-section.
Marked numbers shown in the contour lines correspond to the values of normalized absorption cross section of individual channels:  $2\pi \sigma^{\text{abs}}_{n}/(2n+1) \lambda^2$.
Colored regions are physically allowed solutions; while uncolored regions (or in the White color) represent forbidden solutions.
It is noted that the amplitude square is bounded within the range $[0, 1]$; while the allowed phase is within $[\pi/2, 3\pi/2]$.
The Green cross-marker, localed at ($\theta_n = \pi, \vert S_n \vert^2 = 0.25$), indicates the maximum value, $0.25$, in the normalized absorption cross-section.
 }
\end{figure}

As shown in Fig. 1, even though we do not write down any exact formulas for this phase diagram, the range to support physical value for the amplitude square only exists within $0 \le \vert S_n\vert^2 \le 1$ ; while the phase is bound within  $\pi/2 \le \theta_n \le 3\pi/2$.  
It is worth to remark that  this  range to support our parameter space results from the intrinsic power conservation.
We depict the allowed regions in colors.
Outside the allowed regions,  uncolored regions  represent forbidden solutions for a passive scatter.
Moreover, lossless absorption happens in the boundary between the allowed and forbidden regions.
Along this lossless contour, there exists a family of solutions which all result in $\sigma^{\text{abs}} = 0$, but can be supported with different  scattering coefficients  in amplitude and phase.
It is known that with a localized surface plasmon in the subwavelength structure can assist lossless resonance condition~\cite{resonance-1,resonance-3}, which corresponds to the point $\vert S_n\vert=1$ and $\theta_n = \pi$ in our phase diagram.
As for the invisible cloaks~\cite{cloak1,cloak2,cloak3,cloak4,cloak5,cloak6}, one can look for the supported solutions near the bottom of the phase diagram, i.e., $\vert S_n\vert = 0$, for dominant channels.

Once the composited material in a scatterer has intrinsic loss, the supported scattering coefficient moves to reside inside the colored region in this  phase diagram.
For each channel, the maximum value in  the normalized absorption cross section is  $2\pi \sigma^{\text{abs}}_{n}/(2n+1) \lambda^2=1/4$, i.e., the Green cross-marker shown in Fig. 1, corresponding to coherent perfect absorbers ~\cite{perfect1,perfect2,perfect3,perfect4}, but which is also associated with the same amount of light scattered.
The phase and amplitude of scattering coefficients to achieve a maximum absorption power is  $\pi$ and $1/2$, respectively. 
Moreover, along the contour for a constant absorption power, there exist a maximum and a minimum values in the scattering amplitude, both at the  phase $\theta=\pi$.
It implies that one may design a scatterer possessing  the same absorption power,  but with  different scattering signals.
 As for the concept on cloaking a sensor~\cite{cloaksensors1} to exact  more information from the outside (corresponding to light absorption) but with the overall scattering performance being suppressed, in the phase diagram one can find the corresponding solutions located at $\theta=\pi$ (conjugate-matched condition) with the minimum scattering amplitude~\cite{superabsorber1}. 
Even though above illustrations are demonstrated for a single channel, by including multiple channels,  the intrinsic single channel limitation can be broken to generate superscattering or superabsorber phenomena~\cite{superresonance1,superresonance2,superresonance3,superresonance4,superabsorber1,superabsorber2}.

\begin{figure}
\centering
\includegraphics[width=8.2cm]{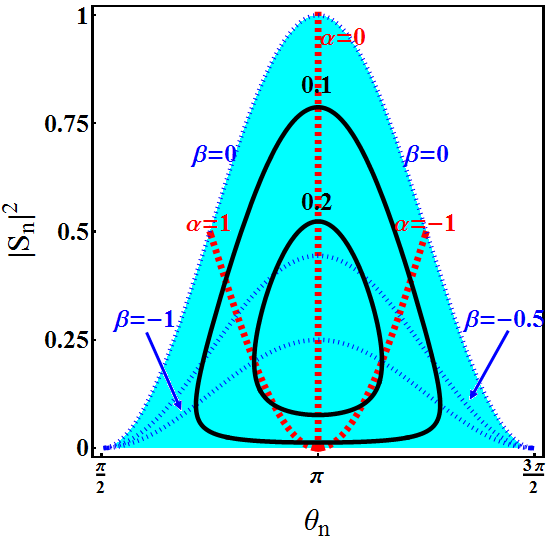}
\caption{(Color online) Supported trajectories in the phase diagram are shown for different sets of the parameters: $\alpha, \beta$ defined in Eq. (3). Here, trajectories with a constant $\beta$ are shown in dotted-curves; while trajectories with a constant $\alpha$ are shown in dashed-curves. Two contours for a constant absorption power are also depicted in the Black color.}
\end{figure}

Through above examples, the phase diagram provides a universal map to all unusual scattering characteristics.
Moreover, through material dispersion or the geometry in particle configurations, one can manipulate the allowed trajectories in the phase diagram to design a passive scatterer with the required scattering and absorption properties.
For this purpose,  we express the scattering coefficient for an isotropic and homogeneous scatterer with multiple layers as 
\begin{eqnarray}
S_{n}&=&-\frac{1}{1+i\frac{V_{n}}{U_{n}}} \equiv  -\frac{1}{1+i(\alpha_{n}+i\beta_{n})}, 
\end{eqnarray}
where $U_{n}$ and $V_{n}$ are determinants of a $2N\times 2N$ matrix constituted by  spherical harmonic functions expressed at each boundary, and $N$ is the number of layers in a  single scatterer ~\cite{resonance-3,cloak1,cloak5}.
Here, we also rewrite this scattering coefficient by introducing two real numbers: $\alpha_{n}$ and $\beta_{n}$, as shown in Eq. (3).
By substituting Eq. (3) into the amplitude square and related phase of scattering coefficients,  one can have different trajectories  in the phase diagram, as shown in Fig. 2.
Supported trajectories for the parameter sets with a constant value of $\alpha$ or $\beta$ are plotted in dashed- and dotted-curves, respectively.

In particular, for the contour with a constant absorption power in the phase diagram, as shown in the Black color in Fig. 2,  one can use the  parametric representation of the curve to describe  this supported trajectory.
The corresponding parameter sets of  $\alpha$ and $\beta$ can be found as
\begin{eqnarray}
\alpha_{n}(t) &=&\sqrt{\frac{1}{4 q_n^2}-\frac{1}{q_n}}\, \sin(t),\\
\beta_{n}(t) &=& (1-\frac{1}{2q_n}) +\sqrt{\frac{1}{4 q_n^2}-\frac{1}{q_n}}\, \cos(t),
\end{eqnarray}
where the independent variable, $t$, is used for the parametric equation, which is also bounded within the range $t=[0,2\pi]$.
Here, for a given normalized absorption power, $q_n \equiv \frac{2\pi}{(2n+1)\lambda^{2}}\sigma_{n}^{\text{abs}}$ in each channel, from Eqs. (4-5)  one can see that the trajectories of $\alpha_{n}$ and $\beta_{n}$ are following an elliptic equation.
In terms of this parametric variable $t$,  in Fig. 3, we show the corresponding absorption and scattering cross-sections for such a contour in the phase diagram. One can see that indeed   the resulting absorption power $\sigma^{\text{abs}}$ is a constant as requested; while there is a degree of freedom to support different scattering powers $\sigma^{\text{scat}}$.
\begin{figure}
\centering
\includegraphics[width=8.0cm]{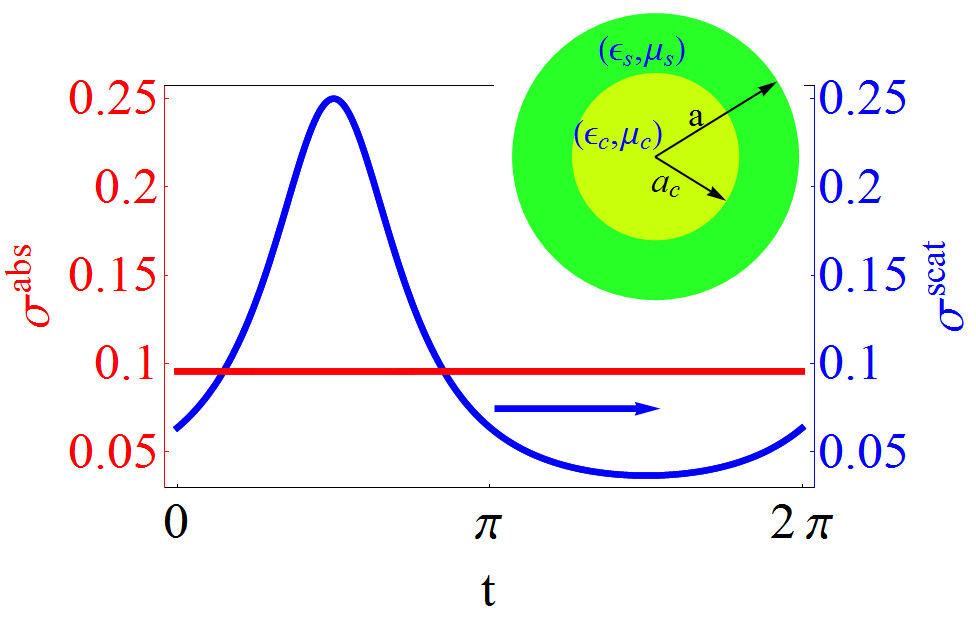}
\caption{(Color online)  Absorption and scattering cross-sections correspond to  the contour shown in Fig. 2, which are depicted in terms of the parametric variable $t$ defined in Eqs. (4-5). Here, the absorption power is a constant as requested,  $\sigma^{\text{abs}} = 0.1$; while there is a degree of freedom in the scattering power $\sigma^{\text{scat}}$.
The insect illustrates the core-shell scatterer used as an  example  to design a passive electromagnetic devices with the constant absorption power.}
\end{figure}
\begin{figure}[h]
\centering
\includegraphics[width=8.2cm]{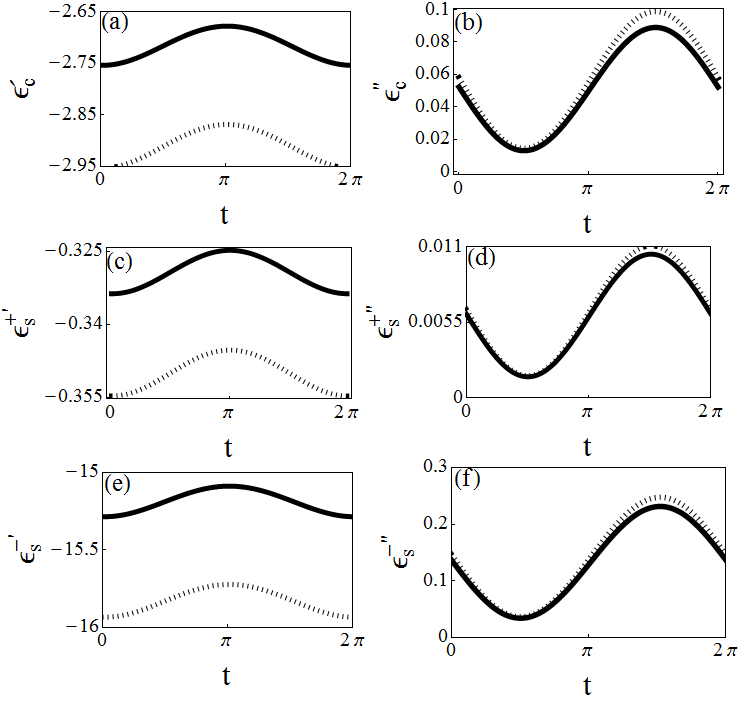}
\caption{(Color online) The permittivities to support a constant absorption power are shown as a function of the parametric variable $t$. For a given material in the shell region, $\epsilon_{s}=3.12$, found solutions for the real and imaginary parts of the permittivity in the core region denoted as $\epsilon_c^\prime$ and $\epsilon_c^{\prime\prime}$ are shown in (a) and (b), respectively.
For a given material in the core region, $\epsilon_{c}=5$, two families of found solutions shown in Eq. (8) are denoted as $\epsilon_s^+$ and $\epsilon_s^-$ for the shell region. The corresponding real and imaginary parts of the permittivity denoted as $\epsilon_s^\prime$ and $\epsilon_s^{\prime\prime}$ are shown in (c, e) and (d, f), respectively.
Results obtained from analytical formulas are depicted in solid-curves; while exact solutions from Mie's theory are depicted in dashed-curves. In all cases, the core-shell geometries are all fixed with $a=1/24\lambda$ and $\gamma=0.9$.}
\end{figure}

In general, the parameter sets $\alpha_{n}$ and $\beta_{n}$ can be manipulated through the geometry, material property, or the working frequency for a scatterer.
As an example, we consider a passive scatterer in the configuration of a core-shell sphere, as illustrated in the insect of Fig. 3, which is composed by two concentric layers of isotropic and homogeneous materials.
The geometrical parameters and material properties for this core-shell scatterer are the radius of core, $a_{c}$, the radius of whole particle, $a$, and $\epsilon_s$($\mu_s$)/ $\epsilon_c$($\mu_c$) for the permittivity (permeability) in the shell/core regions, respectively. 
If the electrically small approximation is satisfied for such a core-shell scatterer, it is known that the main contribution on the corresponding cross-sections dominantly comes from the dipole-wave scattering, i.e., $n = 1$.
The requirement to satisfy is that all the size parameters are small enough, that is $2\pi  a/\lambda\ll 1$, $2\pi  a \sqrt{\epsilon_{s}\mu_{s}} /\lambda \ll 1$, and $2\pi a\sqrt{\epsilon_{c}\mu_{c}}/\lambda \ll 1$ for the outside, shell, and core regions, respectively.
For such a two-layered scatterer, the corresponding scattering coefficients are conducted from a $4\times4$ matrix by tracking TE and TM modes.
However,  for the electric dipole, the TM mode ($n=1$) is dominant. By applying the continuity of electric magnetic fields  established at the each boundary of shell-environment and core-shell, one can approximately express the coefficient $V_{1}^{\text{TM}}/U_{1}^{\text{TM}}$ as
\begin{equation}\label{analytic3}
\frac{V_{1}^{TM}}{U_{1}^{TM}} =\frac{3\lambda^3}{2 (2\pi a)^3}\frac{2\gamma^{3}(1-\epsilon_{s})(\epsilon_{c}-\epsilon_{s})-(2+\epsilon_{s})(\epsilon_{c}+2\epsilon_{s})}{\gamma^{3}(\epsilon_{s}-\epsilon_{c})(2\epsilon_{s}+1)+(1-\epsilon_{s})(\epsilon_{c}+2\epsilon_{s})},\nonumber
\end{equation}
where $\gamma$ is defined as the ratio between the core radius to the whole particle radius, $\gamma \equiv a_{c}/a$.
If one replaces $\epsilon$ by $\mu$, then we can obtain the other coefficient $V_{1}^{\text{TE}}/U_{1}^{\text{TE}}$, which for non-magnetic media is automatically zero for $\mu_{0}=\mu_{s}=\mu_{c}=1$.
By taking $\gamma=1$ or $\epsilon_{s}=\epsilon_{c}$, result shown above can be reduced to the electric dipole equation for a solid sphere.
We want to mark that even though the electrically small approximation is applied here, however, direct numerical results obtained by solving all the infinite channels only give a derivation around $10\%$.

Now, for our core-shell system with the geometric size fixed, we provide a systematic way to find the corresponding material properties with a constant absorption power, as specified by the contour in the phase diagram shown in Fig. 2.
To give a clear illustration, one may fix the material property in the shell or in the core region first.
If we assume that the composition for the shell region is given, i.e., $\epsilon_s$ is fixed, 
Then, based on Eqs. (4-6), the corresponding solution for the permittivity in the core region is
\begin{widetext} 
\begin{equation}\label{analytic1}
\epsilon_{c}=\epsilon_{s}\,\frac{3(-2\epsilon_{s}-4-2\gamma^{3}+2\gamma^{3}\epsilon_{s})-2(\alpha_1 + \beta_1)(2\pi a/\lambda)^{3}(2-2\epsilon_{s}+2\epsilon_{s}\gamma^{3}+\gamma^{3})}{3(\epsilon_{s}+2+2\gamma^{3}\epsilon_{s}-2\gamma^{3})+2(\alpha_1 + \beta_1)(2\pi a/\lambda)^{3}(1-\epsilon_{s}-2\epsilon_{s}\gamma^{3}-\gamma^{3})}.
\end{equation}
\end{widetext}
Solutions obtained from the analytical results shown in  Eq. (7) are shown in Fig. 4(a) and 4(b) for the real and imaginary parts of the  permittivity in the core region, respectively.
In terms of the parametric variable, $t$, we can have a wide rang in selecting materials, and all of them have the same absorption power.
A comparison between Fig. 3 and Figs. 4(a-b), we find that when $t=\pi/2$ the  scattering power reaches a maximum value; while the  $\epsilon_{c}^{\prime\prime}$ for the required material has a minimum value.
The reason comes from that dissipative loss is proportional to $\epsilon_{c}^{\prime\prime}\vert  \vec{E}\vert^{2}$,  where $\vec{E}$ is the local electric field.
In this scenario, with the help of a strong electric field, the loss to maintain the same absorption power can be minimized simultaneously.
In Fig. 4, we also add the numerical calculations obtained by the exact solution from Mie's theory, shown in the dotted-curves accordingly.
A good agreement between our analytical solutions to the exactly numerical ones is demonstrated.

On other hand, if the material property in the core region is specified,  based on Eqs. (4-6) the corresponding solutions for the permittivity in the shell region $\epsilon_s$  to support a constant absorption power are governed by 
\begin{equation}\label{analytic2}
\epsilon_{s}^{\pm}=\frac{-g\pm\sqrt{g^{2}-4fh}}{2f},
\end{equation}
with the shorthanded notations:
\begin{eqnarray}
f&=&2\,(1-\gamma^{3})[3-2 (\alpha_1 + \beta_1)(2\pi a/\lambda)^{3}],\\
g&=&2\,(\alpha_1 + \beta_1)(2\pi a/\lambda)^{3}[\gamma^{3}(1-2\epsilon_{c})+2-\epsilon_{c}]\\ \nonumber
&&+ 3\,(2\gamma^{3}+2\gamma^{3}\epsilon_{c}+\epsilon_{c}+4),\\
h&=&\epsilon_{c}(1-\gamma^{3})[6+2\,(\alpha_1 + \beta_1)(2\pi a/\lambda)^{3}].
\end{eqnarray}
There exists two families to support the materials in the shell region, denoted as $\epsilon_s^\pm$.
Again, we show the real and imaginary parts of the  permittivity in the shell region for these two families in Fig. 4(c-d) and 4(e-f),  as well as the exact solutions in dashed-curves, respectively.
To one's surprise, there exists a variety of choices for the material properties even some specific light scattering or absorption properties is given in the beginning.
We want to emphasize that without the introduction of the phase diagram, it is not only hard to find the required material properties, but also difficult to recognize the power competition and limitation among these cross sections for each channel.

Before the conclusion, we remark that when the electrically small approximation is not valid, multiple channels for TE and TM modes may be excited as expected. 
In this scenario, one can also apply our phase diagram, but not for a single channel only.
Scattering coefficients from several dominant channels are just a natural extension by considering all of them onto the phase diagrams simultaneously.
Through the proper assignment to tune each dominant channels in the  characteristic cross-sections, one may also have unique devices beyond the response of a single channel, such as subwavelength superscattering and superabsorber~\cite{superresonance1,superresonance2,superresonance3,superresonance4,superabsorber1,superabsorber2}.

In summary, in terms of the amplitude square and phase of the scattering coefficients, we introduce a universal phase diagram as a compact tool to link the scattering and absorption powers for each individual channel.
Intrinsically, the power conservation gives the physical boundary in the parameter space for a passive scatterer.
Not only the known exotic scattering and absorption phenomena can be illustrated in this diagram, supported trajectories are also demonstrated to design extrinsic-field-controllable scatterers.
With the core-shell scatterers as an example, we reveal a systematic way to find a variety of solutions in the composited materials, all of which possess the same absorption power.
With the analogy among wave phenomena, the concept of our method can be ready applied to acoustic systems as well as quantum scattering system.

\end{document}